\documentclass[pra,twocolumn,showpacs,preprintnumbers,superscriptaddress]{revtex4}
\usepackage{times}
\usepackage{bm}
\usepackage{graphicx}
\usepackage{amsbsy}
\usepackage{amsmath}
\usepackage{amsfonts}
\usepackage{amsthm}
\usepackage{xcolor}
\usepackage{tikz}
\usetikzlibrary{shapes, arrows}

\begin{document}
	\theoremstyle{plain}
	\newtheorem{theorem}{Theorem}
	\newtheorem{lemma}[theorem]{Lemma}
	\newtheorem{corollary}[theorem]{Corollary}
	\newtheorem{proposition}[theorem]{Proposition}\newtheorem{conjecture}[theorem]{Conjecture}
	\theoremstyle{definition}
	\newtheorem{definition}[theorem]{Definition}
	\theoremstyle{remark}
	\newtheorem*{remark}{Remark}
	\newtheorem{example}{Example}
	\title{Theoretical proposal for the experimental realization of realignment operation}
	\author{Shruti Aggarwal$^1$, Satyabrata Adhikari}
	\email{shruti_phd2k19@dtu.ac.in, satyabrata@dtu.ac.in} \affiliation{Delhi Technological University, Shahbad Daulatpur, Main Bawana Road, Delhi-110042,
		India}
\begin{abstract}
Realignment operation has a significant role in detecting bound as well as free entanglement. Just like partial transposition, it is also based on permutations of the matrix elements. However, the physical implementation of realignment operation is not known yet. In this letter, we address the problem of experimental realization of realignment operation and to achieve this aim, we propose a theoretical proposal for the same. We first show that after applying the realignment operation on a bipartite state, the resulting matrix  can be expressed  in terms of the partial transposition operation along with column interchange operations. We observed that these column interchange operations forms a permutation matrix which can be implemented via SWAP operator acting on the density matrix. This mathematical framework is used to exactly determine the first moment of the realignment matrix experimentally. This has been done by showing that the first moment of the realignment matrix can be expressed as the expectation value of a SWAP operator which indicates the possibility of its measurement. Further, we have provided an estimation of the higher order realigned moments in terms of the first realigned moment and thus pave a way to estimate the higher order moments experimentally. Next, we develop moments based entanglement detection criteria that detect positive partial transpose entangled states (PPTES) as well as negative partial transpose entangled states (NPTES). Moreover, we define a new matrix realignment operation for three-qubit states and have devised an entanglement criteria that is able to detect three-qubit fully entangled states. We have developed various methods and techniques in the detection of bipartite and tripartite entangled states that may be realized in the current technology.
\end{abstract}
	 \maketitle
\section{Introduction}
Entanglement is a fundamental property of quantum systems, promising to power the near future of quantum information processing and quantum computing \cite{niel,guhne2009,rpmkhoro}. Entangled states reveal some peculiar properties of the quantum world, and many interesting quantum information processing tasks can be done better with entangled states than with separable states alone \cite{pshor, bennet93,harrow}.
Determining whether a state is entangled or not is an NP hard problem \cite{gurvit}. Hence, it is important to devise implementable and efficient methods to detect bipartite as well as multipartite entanglement. To solve the entanglement detection problems, researchers have developed many methods such as postive partial transposition (PPT) criteria \cite{peres, horo1998} and computable cross norm or realignment (CCNR) criteria \cite{chenwu}, range criterion \cite{horo1997}, majorization criterion \cite{nielson2001}, construction of positive but not completely positive map e.g. reduction map \cite{horo1999}. Although the detection of entanglement through the construction of witness operator \cite{bruss2002} is well received by the community because it can be realized in the experiment but the PPT criterion and realignment criterion can be considered as most important entanglement detection criterion.
However, realignment criteria is more powerful and operational criteria in the sense that it can detect bound entangled states (PPTES) as well as NPT entangled states \cite{rudolphqip, rudolph2000, rudolph2004}, while PPT criterion detect only NPTES. Since partial transposition is an unphysical operation, it has been approximated to a completely positive map using the method of Structural physical approximation (SPA) \cite{ekert2002}. Like partial transposition, realignment operation is also an unphysical operation and thus the method of SPA of realignment operation has been introduced \cite{shruti3} that successfully detects bound as well as NPT entangled states.\\ 
Both the unphysical operations viz. partial transposition and realignment operation, are based on the permutation of the elements of density matrix \cite{horo2006}. On one hand, the explicit form of partial transposition has been studied extensively due to which this operation has been exploited to a great extent for detection, quantification and characterization of entanglement but on the other hand, the physical interpretation of the realignment operation is still not known which makes the study of this matrix operation more intruiging. Thus, this may be the possible reason that the permutation involved in the realignment operation has not been studied much. Some investigation shows that the realignment operation is a global operation acting on a bipartite state unlike the partial transposition that acts on a subsystem. This makes the physical implementation of realignment operation challenging and could be another possible reason for it to be a comparatively less explored density matrix operation.\\
In the past decades, it has been shown that the moments of the density matrix contains much information about the state and measurement of these moments are relatively easy \cite{sjvan,njohn}. To avoid the state tomography, several moment based methods have been proposed in the literature \cite{elben,neven,imai,liu22,shruti4}. Moments act as practical tools in estimating some crucial properties of quantum systems. A method for experimentally measuring the moments by measuring the purity of state is given in \cite{elben19, brydges}.  A scheme based on the random unitary evolution and local measurements on single-copy quantum states given in \cite{yzhou} is shown to be more practical compared with former methods based on collective measurements on many copies of the identical state. A method to measure the $k$ partial moments using $k$ copies of the state and SWAP operators have been discussed in \cite{alves,sugoto,carteret,wong}. However, measurement of moments of the realignment matrix may be considered as an open problem. This has provided a strong motivation to develop an experimentally feasible method to estimate the moments of the realigned matrix.\\
In this work, our aim is to study the physical nature of permutations involved in the realignment operation and devise physically implementable entanglement detection criteria based on realignment operation. Researchers have shown that the moments of the realigned matrix can be used to detect bound entangled states and hence estimation of these moments is an important task \cite{tzhang, shruti4, liu22}. However, it is not known yet how to estimate the realigned moments so we have considered this problem and shown that these are expressible in terms of expectation value of partial transposition of a permutation operator with respect to any $d\otimes d$ dimensional bipartite quantum state. We have made an effort to explore a method that enables the experimental realization of realignment operator. We have achieved this by constructing the realignment matrix with the help of SWAP operator and partial transposition operation. We have shown that the first moment of the realigned matrix may be determined exactly in an experiment. Also, it may be used in an experiment to detect and estimate the amount of entanglement in a $d\otimes d$ dimensional system. We also provided a method that estimates the $k$th moment of the hermitian matrix $(\rho^R)^{\dagger} \rho^R$ using its first moment. This helps in deriving the entanglement detection criterion in $m\otimes n$ dimensional system that may be experimentally feasible. Moreover, we have generalised the matrix realignment operation in three-qubit system. We will show that the newly defined realignment operator along with the structural physical approximation of partial transposition (SPA-PT) map may be used to develop an entanglement detection criteria that has potential to detect three qubit fully entangled states.\\ 
The paper is organized as follows. In Sec. II, we discuss a few mathematical results that will be needed in the later section. In Sec. III, we review the realignment operation that existed in the literature. In Sec. IV, we explain the procedure for the practical estimation of the first moment of realigned matrix. In Sec. V, we discuss the detection and quantification of $d \otimes d$ dimensional bipartite entangled state using the first moment of the realigned matrix. In Sec. VI, we estimate the kth moments of the hermitian matrix $(\rho_{AB}^R)^{\dagger}\rho_{AB}^R$ by using the first moment of the realigned matrix. In Sec. VII, we generalize the idea of the realignment operation on three-qubit system and derive the criterion for the detection of three-qubit fully entangled states. Finally we conculde in Sec. VIII.
\section{A Few mathematical results}
Let us consider a matrix $A$ of order $m \times n$ with  singular values $\sigma_{i},i=1,2,...n$ arranged in the increasing order $\sigma_{max}=\sigma_{1}\geq \sigma_{2} \geq \sigma_{3}\geq....\geq \sigma_{n}=\sigma_{min}$. To start with, let us recall the definition of trace norm and Frobenius norm.\\
(i) \textbf{Trace norm:} The trace norm $||A||_1$ of an $m \times n$ matrix $A$ is defined as the  sum of singular values of $A$
\begin{eqnarray}
	||A||_1 = \sum_{i=1}^{n} \sigma_{i}(A)
	\label{trnorm}
\end{eqnarray}
(ii) \textbf{Frobenius norm:} The norm $||A||_2$ is called Hilbert-Schmidt norm or Frobenius norm of $A$ and is equal to the sum of the squares of the singular values of $A$.
\begin{eqnarray}
	||A||_2^{2} = \sum_{i=1}^{n} \sigma_{i}^{2}(A)
	\label{frob}
\end{eqnarray}
We now state some mathematical results that we will use later to prove our results and theorems.\\
\textbf{Result 1:} For any matrix $A \in M_n(\mathbb{C})$, we have
\begin{eqnarray}
	|Tr(A)| \leq ||A||_1 
	\label{frob}
\end{eqnarray}
\textbf{Result 2:} \textit{Weyl's Inequality \cite{horn}:} Let $A$ and $B$ be Hermitian matrices in $M_n(\mathbb{C})$. Then the following inequality holds
\begin{eqnarray}
	\lambda_{min}(A) + \lambda_{min}(B) \leq \lambda_{min}(A+B) \label{weyl}
\end{eqnarray}
where $\lambda_{min}(.)$ denotes the minimum eigenvalue of the corresponding matrix.\\
\textbf{Result 3:} For any two $n \times n$ hermitian matrices $A$ and $B$, the following inequality holds \cite{jb}
\begin{eqnarray}
	\lambda_{min}(A) Tr[B] \leq Tr[AB] \leq \lambda_{max}(A) Tr[B] \label{jb}
\end{eqnarray}
\textbf{Result 4:} For any matrix $A \in M_n(\mathbb{C})$ with rank $k$, we have \cite{zou}
\begin{eqnarray}
	||A||_1 \leq \sqrt{k} ||A||_2  \label{normineq}
\end{eqnarray}
where $||A||_1 = Tr[\sqrt{A^{\dagger}A}]$ denotes the trace norm and $||A||_2 = \sqrt{Tr[A^{\dagger}A]}$ denotes the Frobenius norm.
\section{Realignment operation in bipartite system}
\noindent Let us start with the definition of the realignment operation for bipartite system. Interestingly, the realignment operation can be defined in different ways that are equilvalent to each other \cite{rud2003}. Consider a quantum state described by a density matrix $\rho_{AB}$ in $m \otimes n$ dimensional system. It can be constructed in block matrix form with $m$ number of blocks in each row and column with each block $Z_{ij}$ of size $n \times n$. Then the matrix of order $m^2 \times n^2$ obatined after applying the realignment operation on $\rho_{AB}$ may be defined as
\begin{eqnarray}
	\rho_{AB}^R = \begin{pmatrix}
		vec(Z_{11})\\
		.\\
		.\\
		vec(Z_{1n})\\
		.\\
		.\\
		vec(Z_{n1})\\
		.\\
		.\\
		vec(Z_{mn})
	\end{pmatrix} \label{rdef}
\end{eqnarray}
where for any $m\times n$ matrix $X = (x_{ij})$, the vector $vec(X)$ is defined as
\begin{eqnarray}
	vec(X)=	(x_{11}, ., ., ., x_{1n}, x_{21}, ., ., ., x_{2n}, ., ., ., x_{m1}, ., ., ., x_{mn} ) \label{vecx}
\end{eqnarray}
To illustrate, consider the following bipartite qubit state $\rho_{AB}$ in $\mathbb{C}^2 \otimes \mathbb{C}^2$ dimensional system
\begin{eqnarray}
	\rho_{AB}=
	\begin{pmatrix}
		\rho_{11} & \rho_{12} & \rho_{13} & \rho_{14}\\
		\rho_{12}^* & \rho_{22} & \rho_{23} & \rho_{24}\\
		\rho_{13}^* & \rho_{23}^* & \rho_{33} & \rho_{34}\\
		\rho_{14}^* & \rho_{24}^* & \rho_{34}^* & \rho_{44}\\
	\end{pmatrix}
\label{rho}
\end{eqnarray}
where $*$ represents the conjugate.
The matrix obtained after applying the realignment operation $R$  defined in (\ref{rdef}) on the state $\rho_{AB}$ is given as
\begin{eqnarray}
	\rho_{AB}^R = 
	\begin{pmatrix}
	\rho_{11} & \rho_{12} & \rho_{12}^* & \rho_{22}\\
	\rho_{13} & \rho_{14} & \rho_{23} & \rho_{24}\\
	\rho_{13}^* & \rho_{23}^* & \rho_{14}^* & \rho_{24}^*\\
	\rho_{33} & \rho_{34} & \rho_{34}^* & \rho_{44}\\
	\end{pmatrix}
\end{eqnarray}
Realignment operation acting on the bipartite system may be used to detect whether the given bipartite state is entangled or not. The criterion developed for the detection of entanglement in bipartite system is known as CCNR criteria or realignment criteria \cite{chenwu,rudolph2004}. It provides a necessary condition for separability of a quantum state i.e. if a bipartite state $\rho_{AB}$ is separable then $||\rho_{AB}^R||_1 \leq 1$ where $||.||_1$ defines the trace norm. As realignment criterion is only a necessary condition so the entanglement may be detected by the following statement: If the bipartite quantum state satisfies $||\rho_{AB}^R||_1 > 1$ then the state $\rho_{AB}$ is an entangled state. 
\section{Estimation of First moment of realigned matrix}
\noindent It is known that measuring the moments of the density matrix is practically possible using $m$ copies of the state and controlled swap operations \cite{carteret}. Exploiting the similar approach, Gray et al. have shown that the moments of the partially transposed density matrix can also be measured using swap operators \cite{sugoto}. For a state $\rho_{AB}$, it has been shown that the $k$th order moment of the partial transposed matrix $\rho_{AB}^{T_{B}}$  can be measured using the individual constituents of the $k$ copies of the state, i.e., $\rho_{AB}^{\otimes k}= \otimes_{c=1}^k \rho_{A_cB_c}$ \cite{sugoto}. The idea is to write the matrix power as an expectation of a partial transposition of a permutation operator with respect to the state $\rho_{AB}$. This may be considered as an important step since partial transposition operation is not a physical operation. However, no such protocol has been discussed in the literature till now for measurement of the moments of the realignment matrix.\\		
In this section, we employ a similar approach to show that the first moment of $\rho_{AB}^R$ i.e., $Tr[\rho_{AB}^R]$ can be expressed as the expectation value of the partial transposition of a permutation operator with respect to the state $\rho_{AB}$ and thus can be experimentally measurable. Later we derive a few results based on the first moment of $\rho_{AB}^R$. Finally we present a separability criteria and show that it is equivalent to the original realignment criteria for a known class of states. 
\subsection{Realigned matrix as the product of swap operator and partial transposition operation}
In the literature, it has been found that realignment operation may be defined as a way of arranging the elements of the matrix in block matrix form, which is different from the partial transposition operation. We will show here that the realigned matrix of any $m\otimes n$ dimensional bipartite state $\rho_{AB}$ may be obtained by the following two actions: (i) interchange the two columns of the density matrix $\rho_{AB}$ and (ii) apply the partial transposition operation on a subsystem. This may be illustrated by the following figure (Fig.1).
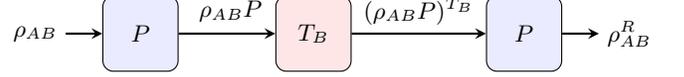
\begin{figure}[h!]
	\begin{tikzpicture}[node distance=1.5cm]
		\tikzstyle{rho} =[]
		\node (in0) [rho] {$\rho_{AB}$};
		\tikzstyle{P} = [rectangle, rounded corners, minimum width=1cm, minimum height=1cm,text centered,  draw=black, fill=blue!8]
		\node (start) [P, xshift=1.4cm] {$P$};
		
		\tikzstyle{Ptb} = [rectangle, rounded corners, minimum width=1cm, minimum height=1cm,text centered, draw=black, fill=red!10]
		\node (in1) [Ptb, xshift=3.7cm] {$T_B$};
		
		\tikzstyle{P2} = [rectangle, rounded corners, minimum width=1cm, minimum height=1cm,text centered, draw=black, fill=blue!8]
		\node (in2) [P2, xshift=6.5cm] {$P$};
		
		\tikzstyle{rhor} =[]
		\node (in3) [rhor, xshift=7.9cm] {$\rho_{AB}^R$};
		
		\tikzstyle{mid1} =[]
		\node (in4) [mid1, xshift=2.6cm, yshift=0.3cm] {$\rho_{AB}P$};
		
		\tikzstyle{mid2} =[]
		\node (in5) [mid2, xshift=5.1cm, yshift=0.3cm] {$(\rho_{AB}P)^{T_B}$};
		
		\tikzstyle{arrow} = [thick,->,>=stealth]
		\draw [arrow] (in0) -- (start);
		\draw [arrow] (start) -- (in1);
		\draw [arrow] (in1) -- (in2);
		\draw [arrow] (in2) -- (in3);
		<TikZ code>
	\end{tikzpicture}	\label{pic1}
	\caption{Schematic diagram representing the realignment operation in terms of partial transposition and permutation operator}
\end{figure}
To understand these actions, let us recall a two-qubit bipartite state described by the density operator $\rho_{AB}$ given in (\ref{rho}). After the first action, the density matrix $\rho_{AB}$ reduces to
\begin{eqnarray}
\rho_{AB}^{(12)}=
\begin{pmatrix}
\rho_{11} & \rho_{13} & \rho_{12} & \rho_{14}\\
\rho_{12}^* & \rho_{23} & \rho_{13} & \rho_{24}\\
\rho_{13}^* & \rho_{33} & \rho_{23}^* & \rho_{34}\\
\rho_{14}^* & \rho_{34}^* & \rho_{24}^* & \rho_{44}\\
\end{pmatrix}
\label{rho12}
\end{eqnarray}
Applying partial transposition operation on qubit $B$, $\rho_{AB}^{(12)}$ reduces to
\begin{eqnarray}
(\rho_{AB}^{(12)})^{T_{B}}=
\begin{pmatrix}
\rho_{11} & \rho_{12}^* & \rho_{12} & \rho_{13}\\
\rho_{13} & \rho_{23} & \rho_{14} & \rho_{24}\\
\rho_{13}^* & \rho_{14}^* & \rho_{23}^* & \rho_{24}^*\\
\rho_{33} & \rho_{34}^* & \rho_{34} & \rho_{44}\\
\end{pmatrix}
\label{rhopt}
\end{eqnarray}
The transformed matrix $(\rho_{AB}^{(12)})^{T_{B}}$ is nothing but the realigned matrix $\rho_{AB}^{R}$.  Therefore, we have shown that realignment operation, which is a non-physical operation may be expressed in terms of another non-physical operation (i.e. partial transposition operation) together with the permutation operation. It may be observe that the interchange of two columns may be realized in an experiment using a swap operator. Thus, if $\rho_{AB}$ denote a $d \otimes d$ dimensional state and $\rho_{AB}^R$ denotes the matrix obtained after applying realignment operation to $\rho_{AB}$, then the realigned matrix $\rho_{AB}^R$ can be expressed as
\begin{equation}
\rho_{AB}^R = (\rho_{AB} P)^{T_B}P  \label{rhorp}
\end{equation}
where $T_B$ denotes the partial transposition operation with respect to the subsystem $B$ and $P = \sum_{i,j=0}^{d-1} |ij\rangle \langle ji|$ denotes the SWAP operator in the computational basis. The SWAP operator $P$ has the following properties:
\begin{enumerate}
\item $P$ is a unitary operator with $P^2 = I_{d^2}$ where $I_{d^2}$  denotes the identity operator in $d \otimes d$ dimensional system.
\item $Tr[P] = Tr[P^{T_B}] = d$
\item  $PP^{T_B} =  P^{T_B}$ where 
$P^{T_B} = \sum_{i,j=0}^{d-1} |ii\rangle \langle jj|$
denote the partial transposition of the SWAP operator.
\end{enumerate}
	
\subsection{Estimation of the first moment of $\rho^R$}
\noindent We are now in a position to estimate the first moment of the realigned matrix $\rho_{AB}^{R}$, where $\rho_{AB}$ denote the $d\otimes d$ dimensional bipartite system.\\
\textbf{Theorem 1.} Let $\rho_{AB}$ be a $d \otimes d$ dimensional bipartite state. If $\rho_{AB}^R$ denote its realigned matrix then the first moment of $\rho_{AB}^R$ is given by
\begin{eqnarray}
t_1 :=	Tr[\rho_{AB}^R] = Tr[\rho_{AB} P^{T_B}] \label{thm1}
\end{eqnarray}
where $P$ is the SWAP operator defined in (\ref{rhorp}).\\
\textit{Proof:} The first moment of $\rho_{AB}^R$ may be defined as
\begin{eqnarray}
t_1 :=	Tr[\rho_{AB}^R] 
\label{def1}
\end{eqnarray}
Using (\ref{def1}) and the expression of $\rho_{AB}^R$ given in (\ref{rhorp}), we get 
\begin{eqnarray}
t_1 :=	Tr[\rho_{AB}^R] &=& Tr[(\rho_{AB} P)^{T_B}P] \nonumber\\
&=& Tr[(\rho_{AB}P) P^{T_B}] \label{e1}\nonumber\\
&=& Tr[\rho_{AB} (P P^{T_B})]\nonumber\\
&=& Tr[\rho_{AB} P^{T_B}] 
\label{e2}
\end{eqnarray}
The equality in the second line follows from the relation $Tr[X Y^{T_B}] = Tr[X^{T_B} Y]$ and the last step follows from the fact that $P  P^{T_B} =  P^{T_B}$. Hence proved \rule{1.2ex}{1.2ex}\\
Therefore, we may conclude that the first moment of $\rho^R$, i.e., $t_1 = Tr[\rho^R]$ can be measured experimentally, since it can be expressed as the expectation value of the partial transposition of the permutation operator with respect to any $d \otimes d$ dimensional bipartite state $\rho_{AB}$.
\section{Usefulness of the first moment of realigned matrix}	
\noindent In this section, we show that the first moment of the realigned matrix may be used in the detection and quantification of entanglement.
\subsection{Detection of $d \otimes d$ dimensional  entangled state}   	
\noindent The realignment criterion can be considered as strong criterion as it detect both PPTES and NPTES. But it is not known whether this criterion may be implemented in the laboratory or not? To address this question, we show that the realignment criterion may be re-expressed in terms of the first moment of the realigned matrix and thus it may be realized in an experiment.\\ 
\textbf{Corollary 1.} If a bipartite state $\rho_{AB}$ is separable, then 
\begin{eqnarray}
t_{1}:=Tr[\rho_{AB} P^{T_B}] \leq 1
\end{eqnarray}
\textit{Proof:} Let us start with the statement of the realignment criterion. It state that if a bipartite state $\rho_{AB}$ is separable then 
\begin{eqnarray}
	||\rho_{AB}^R||_1 \leq 1
	\label{realigncr}
\end{eqnarray}
where $||.||_1$ defines the trace norm.\\
Using Result 1 and (\ref{realigncr}), we get
\begin{eqnarray}
	Tr[\rho^R] \leq ||\rho^R||_1 \leq 1
	\label{ineq301}
\end{eqnarray} 
From Theorem-1, we have $Tr[\rho P^{T_B}] = Tr[\rho^R]$.
Therefore, the theorem is proved using (\ref{ineq301}) \rule{1.2ex}{1.2ex}\\
Let us consider a class of Schmidt-symmetric states defined in \cite{hertz}. As $Tr(\sigma_{AB}^R) = ||\sigma_{AB}^R||_1$ for all Schmidt-symmetric state $\sigma_{AB}$ so the above criteria will become equivalent to the original realignment criteria. Thus, the result given in Corollary 1 will be strong when considering the class of Schmidt-symmetric states. It can be shown that Corollary 1 is nothing but provide us an alternative way to construct the entanglement witness operator which is given by $\alpha I-|\phi\rangle\langle \phi|$, where $|\phi\rangle=\frac{1}{\sqrt{d}}\sum_{i=0}^{d-1}|ii\rangle$.
\subsection{Quantification of $d \otimes d$ dimensional entangled state}
In case of a bipartite system, concurrence is a well known measure of entanglement. An elegant formula of concurrence for two-qubit states has been given by Wootters \cite{wooters}. However, the closed formula of concurrence in higher dimensional quantum system is not known yet. To overcome this situation, researchers have proposed several analytical lower bounds of concurrence for mixed bipartite states in higher dimension \cite{albevario,shruti1,mintert}. The lower bound of the concurrence has been studied through witness operator in the literature \cite{shruti1,mintert}. Witness operator may be implementable in the experiment but it is not known that the lower bound of the concurrence obtained through witness operator always gives better than the bound obtained by Chen et al. \cite{albevario}. Thus there may exist $d \otimes d$ dimensional bipartite entangled state for which Chen et al. bound is better than the bound obtained through witness operator. Hence it is important to review the lower bound of the concurrence and express it in terms of the first moment of the realigned matrix. If we are able to do this then the lower bound of the concurrence given by Chen et al. \cite{albevario} may be realized in the experiment. To achieve our aim, let us re-write the lower bound of the concurrence for arbitrary $d \otimes d$-dimensional system as\cite{albevario}
\begin{eqnarray}
	C(\rho_{AB}) \geq \sqrt{\frac{2}{d(d - 1)}}\left(max(||\rho_{AB}^{T_B}||_1, ||\rho_{AB}^R||_1)-1\right) \label{albe}
\end{eqnarray}
Using the property of trace norm, we have the following inequality 
\begin{eqnarray}
	Tr(\rho_{AB}^{T_B}P) &\leq& ||\rho_{AB}^{T_B}P||_1 \leq ||\rho_{AB}^{T_B}||_1 ||P||_1 = d^2 ||\rho_{AB}^{T_B}||_1\nonumber\\
	\label{clb1}
\end{eqnarray}
The inequality (\ref{clb1}) may be re-expressed in the simplified form as
\begin{eqnarray}
	||\rho_{AB}^{T_B}||_1 \geq \frac{Tr(\rho_{AB}^{T_B}P)}{d^2}=\frac{Tr(\rho_{AB}P^{T_{B}})}{d^2}
	\label{clb2}
\end{eqnarray}
Thus, we have $ max(||\rho_{AB}^{T_B}||_1, ||\rho_{AB}^R||_1) \geq max(\frac{Tr[\rho_{AB}P^{T_{B}}]}{d^2},Tr[\rho_{AB}P^{T_{B}}]) = Tr[\rho_{AB}P^{T_{B}}]$. Hence, the lower bound of concurrence can be derived as 
\begin{eqnarray}
	C(\rho_{AB}) \geq \sqrt{\frac{2}{d(d - 1)}}[Tr(\rho_{AB}P^{T_B}) - 1]
	\label{lbcon}
\end{eqnarray}
From (\ref{lbcon}), it is clear that the lower bound of the concurrence of any $d \otimes d$ dimensional entangled system $\rho_{AB}$ may be estimated through the first moment of the realigned matrix of $\rho_{AB}$. Since the value of  $Tr(\rho_{AB}^{R})=Tr(\rho_{AB}P^{T_B})$ may be estimated practically so the estimation of the lower bound of the concurrence may be possible in an experiment.  
\section{Estimation of the Moments of the hermitian matrix $(\rho_{AB}^R)^\dagger \rho_{AB}^R$}
We may observe that the results obtained above may be useful to detect $d\otimes d$ entangled system but may not be useful in detecting $m\otimes n$ dimensional entangled states. So to deal with this problem, let us first define the $k$th moment of hermitian positive semi-definite operator $(\rho_{AB}^R)^\dagger \rho_{AB}^R$. It may be defined as
\begin{eqnarray}
	T_k = Tr[((\rho_{AB}^R)^\dagger \rho_{AB}^R)^k]
\end{eqnarray} 
The moments of the operator $(\rho_{AB}^R)^\dagger \rho_{AB}^R$ can be considered for the following reasons: (i) $\rho_{AB}^R$ is not hermitian and (ii) an entanglement detection criteria based on moments of the matrix $(\rho_{AB}^R)^{\dagger} \rho_{AB}^R$  has been proposed in \cite{shruti4} but the results obtained may not be easily accessible to the experimentalist.\\
In this section, firstly we derive a lower bound of the first moment of  $((\rho_{AB}^R)^\dagger \rho_{AB}^R)$ in terms of the first moment of $\rho_{AB}^R$ that has been shown to be experimentally measurable. Then we estimate the $k$th moment of $((\rho_{AB}^R)^\dagger \rho_{AB}^R)$ that may also be expressed as the product of the maximum singular value of $\rho_{AB}^{R}$ and the first moment of $((\rho_{AB}^R)^\dagger \rho_{AB}^R)$. Finally, we present an entanglement detection criteria based on these moments and a few examples are given to illustrate that the derived criteria is able to detect bound as well as NPT entangled states.
\subsection{Estimation of the first moment and kth moment of $(\rho_{AB}^R)^\dagger \rho_{AB}^R$ }
Let us consider a $m\otimes n$ dimensional quantum state described by the density operator $\rho_{AB}$ and the realigned matrix of it is denoted by $\rho_{AB}^R$. Assume that $\rho_{AB}^R$ has $k$ non-zero singular values that may be arranged in an assending order as $\sigma_1 \geq \sigma_2 \geq . . . \geq \sigma_k$, where $1\leq k \leq min\{m^2,n^2\}$.\\ 
\textbf{Lemma 1:} The first moment $T_{1}$ of the hermitian positive semidefinite matrix $(\rho_{AB}^R)^\dagger \rho_{AB}^R$ may be bounded below by the following bound:
\begin{eqnarray}
T_1 \geq \frac{(Tr[\rho_{AB} P^{T_B}])^2}{k} \label{T1lb}
\end{eqnarray}
where $k$ denotes the number of non zero singular values of $\rho_{AB}^R$.\\
\textit{Proof:} Applying Result 1 and using $Tr[\rho^{R}]=Tr[\rho^{T_{B}}P]$, we have
\begin{eqnarray}
	Tr[\rho^{T_B} P] = Tr[\rho^R] \leq ||\rho^R||_1 \label{r11}
\end{eqnarray}
Using Result 4 with $A=\rho^{R}$, we have
\begin{eqnarray}
	||\rho^R||_1 \leq \sqrt{k}||\rho^R||_2 = \sqrt{k \sum_{i=1}^k \sigma_i^2 (\rho^R)}= \sqrt{kT_1} \label{r12}
\end{eqnarray}
where $\sigma_i$ denotes the $ith$ singular value of $\rho^R$ and $T_{1}=\sum_{i=1}^{k}\sigma_{i}^{2}(\rho^{R})$.\\
From (\ref{r11}) and (\ref{r12}), we have
\begin{eqnarray}
	Tr[\rho^{T_B} P] \leq \sqrt{kT_1}
	\label{ineq42}
\end{eqnarray}
Simplifying (\ref{ineq42}), we get the required lemma. Hence proved \rule{1.2ex}{1.2ex}\\
Now, we are in the position to derive the lower and upper bound of the kth moment of the hermitian matrix $(\rho_{AB}^R)^\dagger \rho_{AB}^R$ in terms of $T_{1}$ and it may be expressed by the following theorem.\\ 
\textbf{Theorem 2.} The lower and upper bound of the $k$th moment of  $(\rho_{AB}^R)^\dagger \rho_{AB}^R$ may be given by
\begin{eqnarray}
(\sigma^2_{min}(\rho_{AB}^R))^{k-1} T_1 \leq T_k \leq 	(\sigma^2_{max}(\rho_{AB}^R))^{k-1} T_1
\end{eqnarray}
\textbf{Proof:} Since $(\rho_{AB}^R)^\dagger \rho_{AB}^R$ is Hermitian so we can apply Result 3 on $(\rho_{AB}^R)^\dagger \rho_{AB}^R$. Therefore, using (\ref{jb}), the lower and upper bound of the second moment of the matrix $(\rho_{AB}^R)^\dagger \rho_{AB}^R$ may be obtained as follows.
\begin{eqnarray}
	&&	\lambda_{min}((\rho_{AB}^R)^{\dagger}\rho_{AB}^R) Tr[(\rho_{AB}^R)^\dagger \rho_{AB}^R]  \leq Tr[((\rho_{AB}^R)^\dagger \rho_{AB}^R)^2] \nonumber\\&&\leq  \lambda_{max}((\rho_{AB}^R)^{\dagger}\rho_{AB}^R) Tr[(\rho_{AB}^R)^\dagger \rho_{AB}^R]
\end{eqnarray}
Since, $\lambda_{i} ((\rho_{AB}^R)^{\dagger}\rho_{AB}^R) = \sigma^2_i (\rho_{AB}^R)$, the above inequalities can be re-expressed as 
\begin{eqnarray}
	\sigma_{min}^2(\rho_{AB}^R) T_1 \leq T_2 \leq  \sigma_{max}^2(\rho_{AB}^R)  T_1 
\end{eqnarray}
where $T_{1}=Tr((\rho_{AB}^R)^{\dagger}\rho_{AB}^R)$.\\
Following the same procedure for the $k$th moment, we deduce that the following inequality holds for all $1 < k \leq n$.
\begin{eqnarray}
	\sigma_{min}^2(\rho_{AB}^R) T_{k-1} \leq T_k \leq  \sigma_{max}^2(\rho_{AB}^R)  T_{k-1} 
\end{eqnarray}
By the principle of mathematical induction, we have
\begin{eqnarray}
	(\sigma^2_{min}(\rho_{AB}^R))^{k-1}  T_1 &\leq&	\sigma^2_{min}(\rho_{AB}^R) T_{k-1} \leq T_k \nonumber\\ &\leq&  \sigma^2_{max}(\rho_{AB}^R)  T_{k-1} \leq (\sigma^2_{max}(\rho_{AB}^R))^{k-1}  T_1 \nonumber\\ \nonumber
\end{eqnarray}
Hence proved \rule{1.2ex}{1.2ex}
\subsection{Entanglement detection criteria for $m \otimes n$ dimensional bipartite system}
In this section, we will derive an entanglement detection criterion for $m \otimes n$ dimensional bipartite system. The required entanglement detection criterion is derived in terms of the inequality that involve the kth moment $T_{k}$. The kth moment $T_{k}$ may be estimated in terms of the function of the first moment $t_{1}$ of the realigned matrix. Since $t_{1}=Tr[\rho_{AB} P^{T_B}]$ is an experimentally realizable quantity so the $m \otimes n$ dimensional entangled state may be detected in an experiment.\\
\textbf{Theorem 3.} Let $\rho_{AB}$ be any bipartite state in $m \otimes n$ dimensional Hilbert space. Consider the $k$ non-zero singular values of the realigned matrix $\rho_{AB}^R$ that may be denoted as $\sigma_1, \sigma_2, \ldots \sigma_k$ with $1\leq k \leq min\{m^2,n^2\}$. If $\rho_{AB}$ is separable then the following inequality holds:
\begin{equation}
\frac{t_{1}^2}{k} \leq	1 - k(k-1) {D_k}^{1/k} \label{thm4}
\end{equation}
where $D_k = \prod_{i=1}^{k} \sigma_i^2(\rho^R)$.\\
\textbf{\textit{Proof:}} Let $\rho_{AB}$ be any arbitrary separable state. The separability criterion based on $R-moment$ is given by \cite{shruti4}
\begin{eqnarray}
k(k - 1) D_k^{1/k} + T_1 \leq 1 \label{rmoment1}
\end{eqnarray}
Therefore, the inequality (\ref{rmoment1}) holds for all separable state $\rho_{AB}$. The quantity $D_k$ can be estimated in terms of moments using the bounds given in Lemma 1 and Theorem 3. The inequality given in (\ref{rmoment1}) can be re-written as
\begin{eqnarray}
T_1 \leq 1 - k(k-1) D_k^{1/k} \label{rmoment}
\end{eqnarray}
Thus, the inequality in (\ref{thm4}) follows using (\ref{T1lb}). Hence proved. \rule{1.2ex}{1.2ex}

\subsection{Examples}
\example Let us consider the following class of $3 \otimes 3$ PPT entangled states \cite{hakye}.
\begin{eqnarray}
\rho_{\epsilon} = \frac{1}{N}
\begin{pmatrix}
1 & 0 & 0&0&1&0&0&0 & 1\\
0 & 1/\epsilon^2 &0&1&0&0&0 &0&0\\
0 & 0&\epsilon^2&0&0&0&1 &0&0\\
0 & 1&0&\epsilon^2&0&0&0 &0&0\\
1 & 0&0&0&1&0 &0 &0&1\\
0 & 0&0&0&0 &1/\epsilon^2&0 &1&0\\
0 & 0&1&0&0&0&1/ \epsilon^2 &0&0\\
0 & 0&0&0&0&1&0 &\epsilon^2&0\\
1 & 0&0&0&1&0&0 &0& 1\\
\end{pmatrix}
\label{rhoep}
\end{eqnarray} 
where $\epsilon>0$, $\epsilon \neq 1$ and $N=3(1+\epsilon^2+\frac{1}{\epsilon^2})$ is the normalization constant.\\
Matrix rank of $\rho_{\epsilon}^R$ is 8. 
Inequality in (\ref{thm4}) is violated for $\epsilon \in [0.622496, 0.780349]$ and $[1.281481, 1.606435]$ and hence the bound entangled states lying in this region are detected by the criteria given in Theorem 3.
\example  Consider the class of NPT entangled states in $3\otimes 3$ dimensional system, which is defined as
\cite{shruti4}
\begin{eqnarray}
\rho_a =
\begin{pmatrix}
\frac{1-a}{2} & 0 & 0&0&0&0&0&0 & \frac{-11}{50}\\
0 & 0&0&0&0&0&0 &0&0\\
0 & 0&0&0&0&0&0 &0&0\\
0 & 0&0&0&0&0&0 &0&0\\
0 & 0&0&0&\frac{1}{2} - a&  \frac{-11}{50} &0 &0&0\\
0 & 0&0&0&  \frac{-11}{50} &a&0 &0&0\\
0 & 0&0&0&0&0&0 &0&0\\
0 & 0&0&0&0&0&0 &0&0\\
\frac{-11}{50} & 0&0&0&0&0&0 &0& \frac{a}{2}\\
\end{pmatrix}
\label{rhoa}
\end{eqnarray} 
where $\frac{1}{50} (25 - \sqrt{141}) \leq a \leq \frac{1}{100}(25 + \sqrt{141})$. The inequality in (\ref{thm4}) is violated for all $a$ lying in the above range. Hence the above state is detected by the separability criteria given in Theorem 3. 
\section{Classification of tripartite qubit system}
Let us start with the discussion of the classification of tripartite entanglement. Acin et al. introduced a scheme based on the construction of witness operator for the classification of mixed three-qubit states into different entanglement classes \cite{acin}. Mixed three-qubit states are classified by generalizing the classification of three-qubit pure states. Two three-qubit states are called SLOCC (stochastic local operation and classical communication) equivalent if they can be obtained from each other under SLOCC, otherwise they are SLOCC inequivalent \cite{vidal00,dutta1}. Three-qubit states are classified into six SLOCC inequivalent classes: fully separable, three biseparable and two genuinely entangled (GHZ and W) states \cite{vidal00}. 
A density matrix $\rho_{ABC}$ on the Hilbert space $\mathcal{H}_A\otimes \mathcal{H}_B \otimes \mathcal{H}_C$ ($\mathcal{H}_i,i=A,B,C$, denotes the Hilbert spaces of dimension 2) is fully separable if it can be written as 
\begin{eqnarray}
	\rho_{ABC} = \sum_i p_i \rho_i^{A} \otimes \rho_i^{B} \otimes \rho_i^{C} \;\text{with}\; p_i \geq 0 \; \text{and}\;\sum p_i = 1 \nonumber
\end{eqnarray}
If a state is not fully separable then the state either belongs to biseparable class or to the class of fully entangled states. The three classes of biseparable states contain only bipartite entanglement between any two of the qubits. For example, the states in class $A|BC$ possess entanglement between the systems $B$ and $C$ and are separable with respect to the subsystem $A$. A three-qubit state is said to be fully entangled if it is entangled with respect to every bipartition of the whole system. As we know that the partial transposition operation is not completely positive so the structural physical approximation of partial transposition (SPA-PT) map for the three-qubit system has been studied to classify its different SLOCC inequivalent classes \cite{kumari1}. Here, we use the SPA-PT map for the detection of fully entangled three-qubit states.\\ 
Let $\rho_{ABC}$ be a three-qubit state and  $\rho_{ABC}^{T_A}$, $\rho_{ABC}^{T_B}$,  $\rho_{ABC}^{T_C}$ denote the partial transposition of $\rho_{ABC}$ with respect to the subsystem $A$, $B$ and $C$ respectively.  Let us now consider the SPA-PT of a single qubit in a three-qubit system. The SPA-PT map on the qubit $X$ ($X=A,B,C)$ of
the three-qubit state $\rho_{ABC}$, may be defined as 
\begin{eqnarray}
	\widetilde{\rho^{T_X}_{ABC}} = \frac{1}{10} (I_2 \otimes I_2 \otimes I_2) + \frac{1}{5}\rho^{T_X}_{ABC},~~X=A,B,C \label{spapt}
\end{eqnarray}
where $I_2$ denotes the $2 \times 2$ identity matrix.
The SPA-PT map defined above is completely positive and hence may be implemented in an experiment \cite{kumari1}.
The statement of the necessary conditions for the biseparability of a three-qubit state is given by \cite{kumari1}.
\begin{enumerate}
	\item[(a)]  If a tripartite state $\rho_{ABC}$ is biseparable in the $A|BC$ cut, then 
	\begin{eqnarray}
		\lambda_{min} (\widetilde{\rho^{T_A}_{ABC}}) \geq \frac{1}{10} \label{Anu1}
	\end{eqnarray}
	\item [(b)] If $\rho_{ABC}$ is biseparable in the $B|AC$ cut, then 
	\begin{eqnarray}
		\lambda_{min} (\widetilde{\rho^{T_B}_{ABC}}) \geq \frac{1}{10} \label{anu2}
	\end{eqnarray}
	\item [(c)] If $\rho_{ABC}$ is biseparable in the $C|AB$ cut, then 
	\begin{eqnarray}
		\lambda_{min} (\widetilde{\rho^{T_C}_{ABC}}) \geq \frac{1}{10} \label{anu3}
	\end{eqnarray}
\end{enumerate}	
It may be noted that the violation of the inequalities (\ref{Anu1}), (\ref{anu2}), (\ref{anu3}) implies that the state $\rho_{ABC}$ is either fully separable or fully entangled.
\subsection{Realignment map on three qubit system}
We now generalize the concept of a realignment map on three party system and thus define a realignment operation on $2 \otimes 2 \otimes 2$ system. Let $\rho_{ABC}$ be a quantum state in $2 \otimes 2 \otimes 2$ dimensional Hilbert space. In block matrix form, $\rho_{ABC}$ can be expressed as
\begin{eqnarray}
	\rho_{ABC} =
	\begin{pmatrix}
		A &B&C&D\\
		B^*&E&F&G\\
		C^*&F^*&H&I\\
		D^*&G^*&I^*&J
	\end{pmatrix}
\end{eqnarray}
where $A, B, C, D, E, F, G, H, I$, and $J$ represent a $2 \times 2$ block matrices.\\
The realignment matrix of $\rho_{ABC}$ may be defined as
\begin{eqnarray}
	\rho_{ABC}^{\mathcal{R}} =
	\begin{pmatrix}
		vec(A) & vec(B)\\
		vec(C) & vec(D)\\
		vec(B^*) & vec(E)\\
		vec(F) & vec(G)\\
		vec(C^*) & vec(F^*)\\
		vec(H) & vec(I)\\
		vec(D^*) & vec(G^*)\\
		vec(I^*) & vec(J)
	\end{pmatrix} 
	\label{rhortri}
\end{eqnarray}
For any matrix $Z=(z_{ij}) \in C^{m\times n}$,  $vec(Z)$ is defined as
\begin{eqnarray}
	vec(Z) = (z_{11}, . . . z_{m1}, z_{12}, . . .,z_{m2}, z_{1n}, . . .,z_{mn})
\end{eqnarray}
Let $Q$ be a permutation operator defined by $Q(e_i) = e_{\pi_i}$ where $\pi$ is a permutation on $S_8$ given by $\pi(1,2,3,4,5,6,7,8) = (1,3,5,7,2,4,6,8)$ and $e_i$ denotes a $8 \times 1$ column vector with $i$th entry equal to 1 and other entries equal to zero. Thus, the permutation operator $Q$ may be expreesed in matrix form as
\begin{eqnarray}
	Q=
	\begin{pmatrix}
		1 & 0 & 0 & 0 & 0 & 0 & 0 & 0\\
		0 & 0 & 1 & 0 & 0 & 0 & 0 &0\\
		0 & 0 & 0 & 0 & 1 & 0 & 0&0\\
		0 & 0 & 0 & 0 & 0 & 0 & 1&0\\
		0 & 1 & 0 & 0 & 0 & 0 & 0 &0\\
		0 & 0 & 0 & 0 & 0 & 1 & 0 & 0\\
		0 & 0 & 0 & 0 & 0 & 0 & 0& 1
	\end{pmatrix}
\end{eqnarray}
Let $X =[X_{ij}]_{i,j=1}^4$ be any $8 \times 8$ matrix written in block matrix form with each block matrix $X_{ij}$ of size 2. Then $\tau$ is the matrix operation defined as $X^{\tau} = [X_{ij}^T]_{i,j=1}^4$ where $T$ denote the usual transpose operation. Then the transformed matrix $\rho_{ABC}^{\mathcal{R}}$ obtained after applying the realignment operation on the state $\rho_{ABC}$ can be expressed as
\begin{eqnarray}
	\rho_{ABC}^{\mathcal{R}} = (\rho_{ABC} Q)^{\tau} \label{rhorq}
\end{eqnarray}
Thus, the realigned matrix $\rho_{ABC}^{\mathcal{R}}$ is expressible in terms of a permutation operator which shows that the properties of the proposed operation may be estimated in an experiment.
\subsection{Detection of entanglement in three qubit system}
Now we use the realignment operation defined in (\ref{rhortri}) to derive an entanglement detection criteria for three-qubit states. To develop this criteria, we use the SPA-PT map defined in (\ref{spapt}). This criteria involves the computation of the minimum eigenvalue of the Hermitian matrix $(\rho_{ABC}^\mathcal{R})^{\dagger}\rho_{ABC}^\mathcal{R}$,where $\mathcal{R}$ is the realignment operation defined in (\ref{rhorq}). Later, we will show that our criteria has potential to detect $2 \otimes 2 \otimes 2$ bound entangled states.\\
\textbf{Theorem 4.} Let $\rho_{ABC}$ be a tripartite state in $2 \otimes 2 \otimes 2$ dimensional Hilbert space.\\
(a) If the state $\rho_{ABC}$ is biseparable in the $A|BC$ cut, then 
\begin{eqnarray}
\lambda_{min}((\rho_{ABC}^\mathcal{R})^{\dagger}\rho_{ABC}^\mathcal{R} + \widetilde{\rho_{ABC}^{T_{A}}}) \geq \lambda_{min}((\rho_{ABC}^\mathcal{R})^{\dagger}\rho_{ABC}^\mathcal{R}) + \frac{1}{10} \nonumber\\ \label{s1}
\end{eqnarray}
(b) If the state $\rho_{ABC}$ is biseparable in the $B|AC$ cut respectively, then 
\begin{eqnarray}
	\lambda_{min}((\rho_{ABC}^\mathcal{R})^{\dagger}\rho_{ABC}^\mathcal{R} + \widetilde{\rho_{ABC}^{T_{B}}}) \geq \lambda_{min}((\rho_{ABC}^\mathcal{R})^{\dagger}\rho_{ABC}^\mathcal{R}) + \frac{1}{10} \nonumber\\ \label{s2}
\end{eqnarray}
(c) If the state $\rho_{ABC}$ is biseparable in the $C|AB$ cut respectively, then 
\begin{eqnarray}
	\lambda_{min}((\rho_{ABC}^\mathcal{R})^{\dagger}\rho_{ABC}^\mathcal{R} + \widetilde{\rho_{ABC}^{T_{C}}}) \geq \lambda_{min}((\rho_{ABC}^\mathcal{R})^{\dagger}\rho_{ABC}^\mathcal{R}) + \frac{1}{10} \nonumber\\ \label{s1}
\end{eqnarray}
where $\widetilde{\rho_{ABC}^{T_X}}$ denotes the SPA-PT with respect to the subsystem $X=A,B,C$.\\
\textit{Proof:} Let $\rho_{ABC}$ be a biseparable state in the $A|BC$ cut. Since $(\rho_{ABC}^R)^{\dagger}\rho_{ABC}^R$ and $\rho_{ABC}^{T_A}$ are Hermitian and using (\ref{weyl}), we have
\begin{eqnarray}
	\lambda_{min}((\rho_{ABC}^R)^{\dagger}\rho_{ABC}^R + \widetilde{\rho_{ABC}^{T_A}}) &\geq& \lambda_{min}((\rho_{ABC}^R)^{\dagger}\rho_{ABC}^R) \nonumber \\&+& \lambda_{min}( \widetilde{\rho_{ABC}^{T_A}}) \nonumber \\ &\geq& \lambda_{min}((\rho_{ABC}^R)^{\dagger}\rho_{ABC}^R)\nonumber \\ &+& \frac{1}{10}
\end{eqnarray}
where the last inequality follows from (\ref{Anu1}). Similarly it can be proved for $X=B$ and $X=C$. This proves Theorem 4. \rule{1.2ex}{1.2ex}\\
\textbf{Corollary 2.}  If a tripartite state $\rho_{ABC}$ is fully separable then following inequality holds with respect to every subsystem $X = A, B, C$
\begin{eqnarray}
	\lambda_{min}((\rho^R)^{\dagger}\rho^R + \widetilde{\rho^{T_X}}) \geq \lambda_{min}((\rho^R)^{\dagger}\rho^R) + \frac{1}{10} \label{s5}
\end{eqnarray}
Violation of the inequality (\ref{s5}) for atleast one subsystem $X=A,B$ or $C$ implies that the state is not fully separable. \\
\textbf{Corollary 3.} A tripartite state $\rho_{ABC}$ is fully entangled if the following inequality holds with respect to every subsystem $X = A, B, C$
	\begin{eqnarray}
		\lambda_{min}((\rho^R)^{\dagger}\rho^R + \widetilde{\rho^{T_X}}) < \lambda_{min}((\rho^R)^{\dagger}\rho^R) + \frac{1}{10} \label{s4}
	\end{eqnarray}
The result stated in Corollary 3 is strong in the sense that it has the capability to detect fully entangled three qubit state. 
\subsection{Example}
Consider the following family of entangled three qubit states constructed from mutually unbiased basis as \cite{jaferi}
\begin{eqnarray}
	\rho_{ABC}(p_1,p_2,p_3) &=& \frac{1}{8} [I_2 \otimes I_2 \otimes I_2 + r_1 \sigma_z \otimes \sigma_z \otimes I_2 \nonumber\\&+& r_2 \sigma_z \otimes I_2 \otimes \sigma_z + r_3 I_2 \otimes \sigma_z \otimes \sigma_z \nonumber \\ &+& r_4 \sigma_x \otimes \sigma_x \otimes \sigma_x +  r_5 \sigma_x \otimes \sigma_y \otimes \sigma_y  \nonumber\\
	&+&  r_6 \sigma_y \otimes \sigma_x \otimes \sigma_y + r_7 \sigma_y \otimes \sigma_y \otimes \sigma_x]\nonumber \\  \label{pstate}
\end{eqnarray}
where $r_1 = r_2 = r_3 = p_1 + p_2 - p_3$, $r_4 = p_1 - p_2 + 3p_3$,  $r_5 = r_6 = r_7 = -p_1 + p_2 + p_3$ with $p_1 + p_2 + 3 p_3 = 1$ and $0\leq p_i \leq 1$ for $i=1,2,3$. 
Using the separability criteria given in Theorem 4, we find that the inequality given in (\ref{s1}) is violated for each subsystem $X=A,B,C$ for the values of $(p_1, p_3)$ shown in Fig-\ref{pstate}a., b., and c.
Thus, by Corollary 3, we conclude that $\rho_{ABC}(p_1,p_2,p_3)$ is three qubit fully entangled. Hence the three qubit entangled state $\rho_{ABC}(p_1,p_2,p_3)$ is detected by the criterion given in Corollary 3.
\begin{figure}[h!]
	2a.	\includegraphics[width=0.36\textwidth]{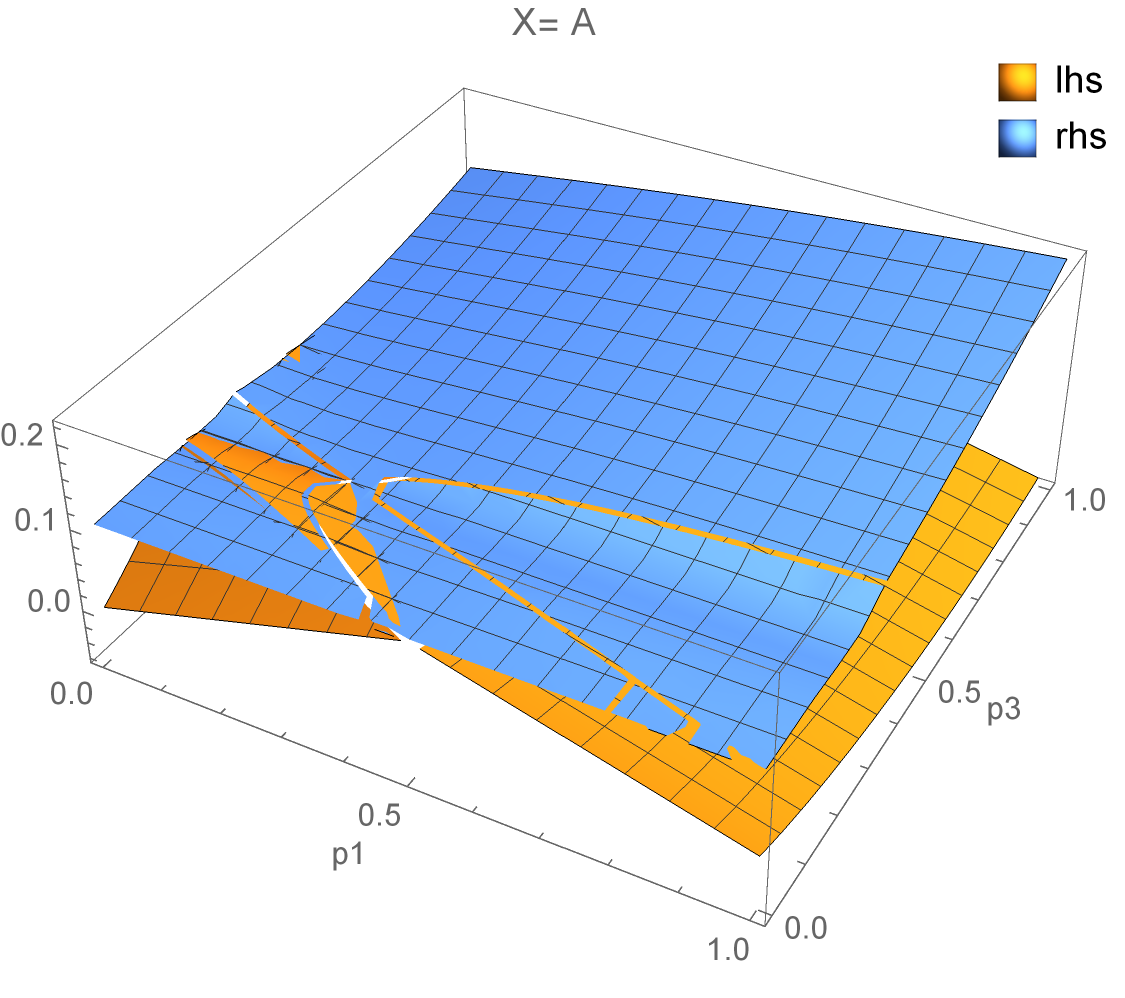}\\
	2b.	\includegraphics[width=0.36\textwidth]{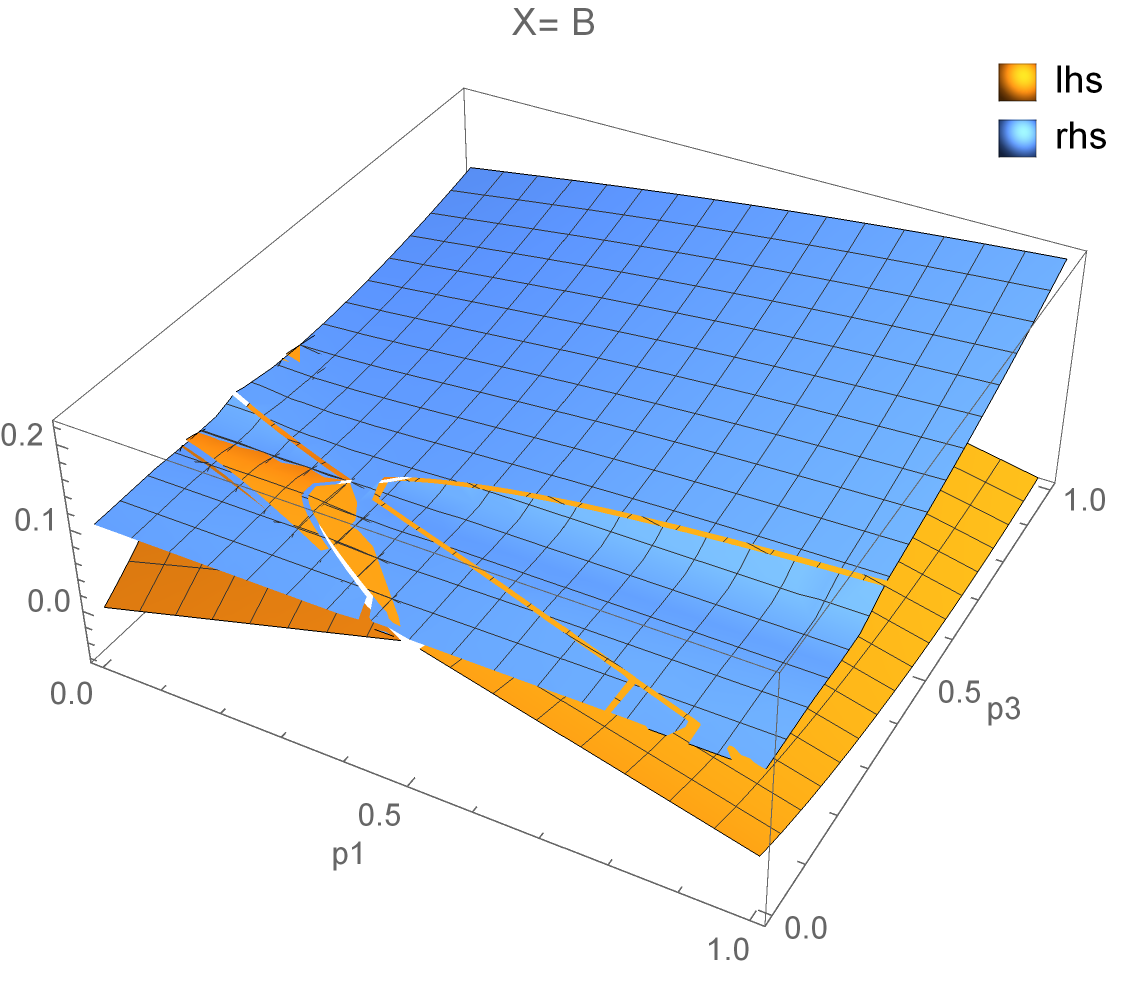}\\
	2c.	\includegraphics[width=0.36\textwidth]{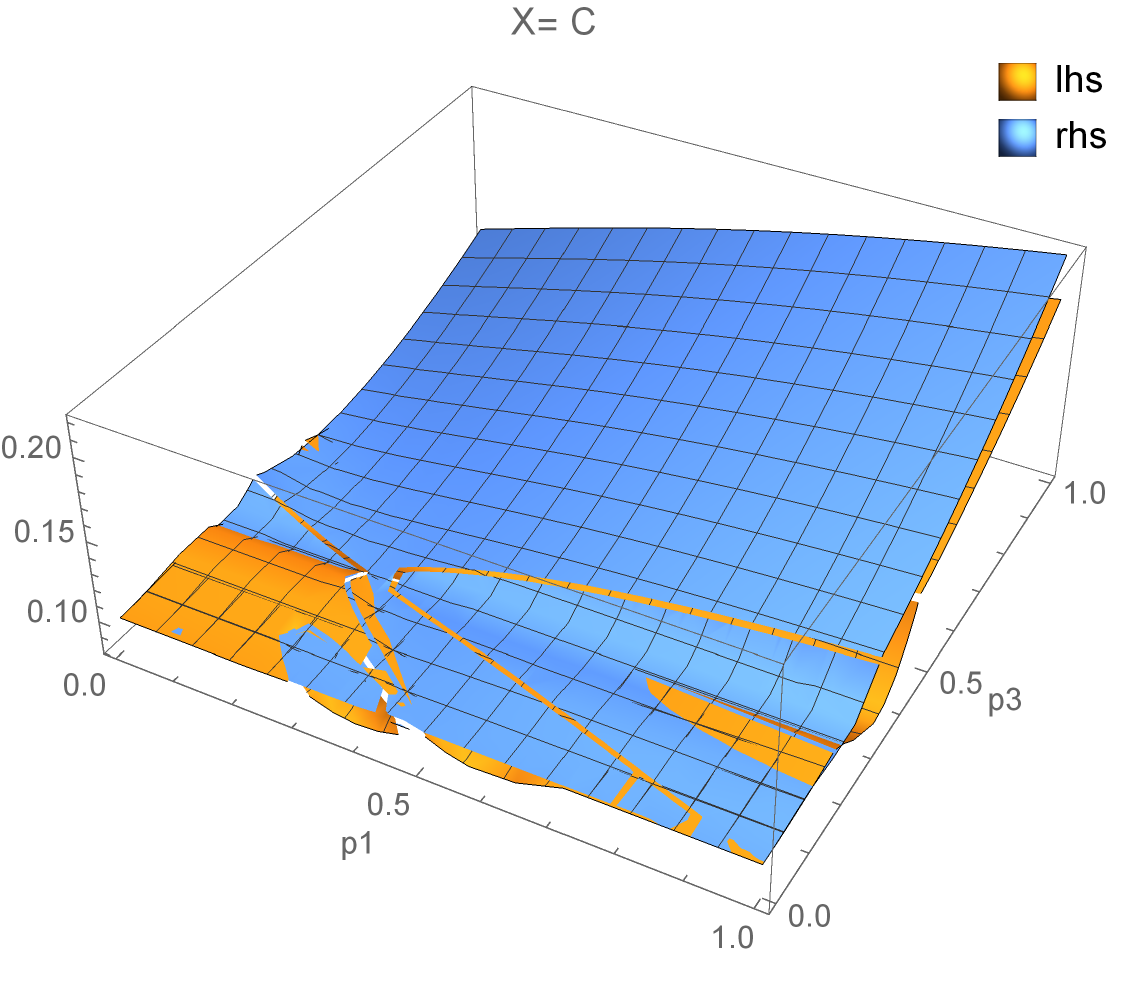}
\caption{For the state $\rho_{p_1,p_2,p_3}$ given in (\ref{pstate}), the yellow and blue curve represents the left hand side and the right hand side of the inequality (\ref{s4}) for $X=A, B$ and $C$ in Fig 2a., 2b., and 2c., respectively. The graph is plotted with respect to the state parameter $p_1$ and $p_3$. The figure shows that the inequality (\ref{s4}) holds for large number of states and hence those entangled states are detected by our method.}
	\label{pstate}
\end{figure}

\section{Conclusion}
To summarize, we have provided an experimentally feasible form of the realignment operation. We have shown that the realignment operation is expressible in terms of partial transposition and permutation operator.
We have demonstrated that the first moment of the realigned matrix can be expressed in terms of the expectation value of SWAP operator which indicates the possibility of its physical measurement. Next, this physically realizable quantity, $Tr[\rho^R]$ has been used to obtain a separability criteria. This criteria is although weaker than the ones in literature in terms of detection power, it is important because it is equivalent to the original realignment criteria for a particular class of states called Schmidt-symmetric states. Since $\rho^R$ is not Hermitian, we have derived the $k$th moments of the Hermitian operator $(\rho^R)^{\dagger} \rho^R$ in terms of the its first moment which reduces the heavy calculations required to compute higher order moments. Using these moments, we propose entanglement detection criteria that are shown to detect bound as well as NPT entangled states. Moreover, we have defined a new realignment operation for three-qubit states. We have shown that the three-qubit realigned matrix can be expressed using a permutation operator. We derive a separability criteria to detect and classify three-qubit fully entangled states using SPA-PT and the proposed realignment operation. It has been shown using an example that the criteria has potential to detect fully entangled states. 

\section{Acknowledgement}
The first author Shruti Aggarwal would like to acknowledge the financial support by Council of Scientific and Industrial Research (CSIR), Government of India (File no. 08/133(0043)/2019-EMR-1).
	
\section{DATA AVAILABILITY STATEMENT}
Data sharing not applicable to this article as no datasets were generated or analysed during the current study.

\end{document}